\documentstyle[aps,epsfig,multicol,prl]{revtex}

\relax    
\begin{document}

\draft

\title{Magneto-thermal instabilities in $\kappa-$(BEDT-TTF)$_2$Cu(NCS)$_2$}

\author{M. M. Mola and S. Hill\cite{email}}
\address{Department of Physics, Montana State University, Bozeman, MT 59717}

\author{J. S. Brooks and J. S. Qualls}
\address{Department of Physics and National High Magnetic Field Laboratory, Florida State University,
Tallahassee, FL 32310}

\date{\today}
\maketitle

\bigskip

\begin{abstract}
Angle and temperature dependent torque magnetization measurements
are reported for the organic superconductor
$\kappa-$(ET)$_2$Cu(NCS)$_2$, at extremely low temperatures
($\sim$T$_c$/$10^3$). Magneto-thermal instabilities are observed
in the form of abrupt magnetization (flux) jumps. We carry out an
analysis of the temperature and field orientation dependence of
these flux jumps based on accepted models for layered type-II
superconductors. Using a simple Bean model, we also find a
critical current density of $4\times10^8$ A.m$^{-2}$ from the
remnant magnetization, in agreement with previous measurements.
\end{abstract}
\bigskip

\pacs{{\em Keywords}:   superconductivity, vortices, critical
state}

\begin{multicols}{2}[]
\centerline{\bf 1. Introduction}
\smallskip

\noindent{Since the discovery of highly anisotropic high$-$T$_c$
superconductors (HTS) and organic superconductors, there has been
a renewed interest in magneto-thermal instabilities within the
mixed superconducting state \cite{1,2,3}. Much of this interest
derives from the limit such instabilities could potentially place
on technological applications. Our recent work has shown that much
of the field-temperature ({\bf B},T) phase diagram for the organic
superconductor $\kappa-$(ET)$_2$Cu(NCS)$_2$ is either within the
quasi-two-dimensional (Q2D) vortex solid phase, or the vortex
liquid state \cite{4}. Thus, we can expect the rigidity of the Q2D
pancake vortex solid to play an important role throughout much of
the available mixed state {\bf B},T phase diagram. To this end, we
have used angle and temperature dependent torque magnetometry to
investigate the mechanism by which magnetic flux enters an extreme
type-II layered organic superconductor, and the instabilities
caused in the process.}

\bigskip
\centerline{\bf 2. Experimental}
\smallskip

A capacitive cantilever torque magnetometer was used to detect
changes in the magnetization of a single crystal of
$\kappa-$(ET)$_2$Cu(NCS)$_2$, with approximate dimensions of
$1.0\times1.0\times0.3$ mm$^3$. The torque beam and the sample
were mounted on a single axis rotator, allowing for angle
dependent measurements. An angle $\theta=0^o$ refers to the
applied field aligned with the least conductive {\em a}-axis,
while $\theta=90^o$ refers to the field parallel to the highly
conducting {\em bc}-planes. Magnetization was obtained by dividing
the measured torque by {\bf H}$_{app}\times\sin\theta$, and the
volume of the sample, where {\bf H}$_{app}$ is the applied field
strength. Measurements were carried out at the National High
Magnetic Field Laboratory using a $^3$He/$^4$He dilution
refrigerator in conjunction with a 20 tesla superconducting
magnet. The applied field was swept at a constant rate of 0.5
tesla per minute.

\bigskip
\centerline{\bf 3. Results and Discussion}
\smallskip

In Fig. 1, we plot magnetization vs. applied magnetic field for
both up and down sweeps (signified with arrows), at 25 mK, and for
angles between $16.2^o$ and $70.1^o$. The magnetization initially
increases on the up sweeps, reaches a maximum, and then decreases
steadily until it reaches zero at the irreversibility field {\bf
H}$_{irr}$; the down sweeps exhibit similar behavior with the sign
of the magnetization inverted. Catastrophic flux jumps are seen as
a series of saw tooth oscillations superimposed upon the steadily
decreasing magnetization curve. Between the flux jumps, the
magnetization increases monotonically with field, with a slope
proportional 1/{\bf H}$_{app}$ (see right inset to Fig. 1). We
note that none of the flux jumps are complete, {\em i.e}. the
magnetization minimum after a given jump never reaches zero. This
implies that the flux gradient and associated critical currents
are never totally dissipated, and the system is always in a
metastable state. Notice also that the limiting magnetization
values fall within a slowly varying (on the scale of the flux
jumps) envelope, consistent with previous measurements on this
material \cite{3} as well as the HTS \cite{1,2}. Moreover,
hysteresis in the jump amplitudes in the up and down sweeps is
observed; the down sweeps show smaller, more frequent flux jumps
than the up sweeps, indicating that flux from a smaller volume of
the crystal is participating in each jump \cite{1}.

In the right inset to Fig. 1, we plot d{\bf M}/d{\bf H}$_{app}$
vs. 1/{\bf H}$_{app}$ for $\theta=74.6^o$. At low values of 1/{\bf
H}$_{app}$, d{\bf M}/d{\bf H}$_{app}$ is small and constant. This
is followed by a sudden increase to a field dependent value, which
increases linearly with respect to 1/{\bf H}$_{app}$. We find this
slope to be a monotonically decreasing function of angle, as can
be seen in the left inset to Fig. 1. The dark, vertical streaks in
the right inset correspond to the derivative taken over the
discontinuities at the flux jumps, which appear as inverted delta
functions. Note also that the average jump amplitude, A($\theta$)
(the difference in magnetization just before and after the
discontinuity), is an increasing function of angle, which
approximately follows a 1 - $\cos(\theta)$ dependence. This
behavior is due to the quasi two dimensional (Q2D) nature of the
material, {\em i.e.} the flux density scales as the cosine of the
angle between the applied field and the direction normal to the
superconducting planes \cite{4}.

Temperature dependent measurements were made between 25 and 200
mK, and at angles of $47.7^o$ and $74.6^o$. Both angles gave
qualitatively similar results, which can then be scaled to
$\theta=0^o$ \cite{4,5}. In Fig. 2, we plot magnetization for both
up and down sweeps, at an angle of $\theta=47.7^o$, and for
temperatures of 87, 110, and 130 mK; each set of sweeps has been
offset for clarity. Notice that the jumps get progressively larger
and that the number of jumps decreases as the temperature
increases.

It has previously been suggested \cite{3} that flux jump behavior
observed at dilution refrigerator temperatures can be attributed
to a thermal Kapitza resistance which isolates the sample from the
surrounding cryogen bath, thereby triggering runaway thermal
instabilities. Due to the Kapitza resistance having a 1/T$^3$
dependence \cite{6} and, therefore, a T$^3$ thermal conductivity,
the cryogen bath is able to remove heat (caused by flux flow) from
the sample more efficiently at higher temperatures. Dissipation
within the crystal arises due to the viscous flow of vortices into
(out of) the sample as {\bf H}$_{app}$ is increased (decreased).
Power dissipation is, therefore, proportional to the square of the
critical screening current density integrated over the volume of
the crystal, {\em i.e}. q(J$_c$) $\propto$ $\int$J$_c^2dV$;
q(J$_c$) also depends on the rate at which flux enters the sample
(the sweep rate), as has been demonstrated in earlier work
\cite{3}. As the temperature is increased, the cryogen bath is
able to remove heat from the sample at a greater rate (power),
which allows a larger integrated current to build up before any
instability occurs. Hence the flux jumps are observed less
frequently, and with greater magnitude, as the temperature is
increased. We note that the magnetization is proportional to the
curl of the critical screening current density integrated over the
volume of the crystal, {\em i.e}. {\bf M} =
$\int(\nabla\times${\bf J}$_c)dV$, thus, {\bf M} $\propto$
q$^{1/2}$. Therefore, it is not surprising that we observe the
amplitude of the flux jumps to scale with temperature as
$\Delta${\bf M} $\propto$ T$^{3/2}$. We believe that the sudden
cessation of the flux jumps above a characteristic field {\bf
B}$_m$(T), which decreases upon increasing the temperature, is due
to a Q2D melting transition (see Ref. \cite{4}).

As seen above, the amount of dissipation depends on the sample
volume through which screening currents flow. Thus, with a greater
capacity for heat removal to the cryogen bath, currents will flow
in a larger volume of the crystal at higher temperatures, causing
more flux to participate in each jump, {\em i.e}. larger, less
frequent instabilities. We also notice that, as the temperature is
increased, the jumps become more complete, indicating that there
is less remnant current within the crystal following each
instability.

Finally, using a simple Bean model \cite{7}, we use the remnant
magnetization to determine the in-plane critical current for
$\kappa-$(ET)$_2$Cu(NCS)$_2$. In Fig. 2, we label the magnetic
field axis at the point where the magnetization equals zero on the
up sweep. This point corresponds to equal numbers of vortices and
anti-vortices within the sample, and is given by {\bf H*} =
$^1/_2$J$_{c\parallel}$d, where d is the diameter of the sample.
Using d $\sim$ 1 mm, and {\bf H*} scaled by the method mentioned
above \cite{5}, we obtain an in-plane critical current density of,
J$_{c\parallel}$ $\sim4\times10^8$ A.m$^{-2}$, in very good
agreement with previous studies using other methods \cite{8}.

\bigskip
\centerline{\bf 4. Conclusions}
\smallskip

We have used torque magnetometry to investigate the temperature
and field orientation dependence of magneto-thermal instabilities
in an organic superconductor. We find that the amplitude of the
observed flux jumps $-$ hence the number of vortices participating
in each jump $-$ is an increasing function of angle. We also find
that the amplitude of the flux jumps increases with temperature as
A $\propto$ T$^{3/2}$, consistent with accepted models. Finally,
we obtain an in-plane critical current density of J$_{c\parallel}$
$\sim4\times10^8$ A.m$^{-2}$, also consistent with previous
measurements.

\bigskip
\centerline{\bf 5. Acknowledgements}
\smallskip

This work was supported by the National Science Foundation
(DMR-0071953), Research Corporation, and the Office of Naval
Research (N00014-98-1-0538).  Work carried out at the NHMFL was
supported by a cooperative agreement between the State of Florida
and the NSF under DMR-9527035.




\end{multicols}
\bigskip
\noindent{{\bf Figures}}
\bigskip
\begin{figure} \centerline{\epsfig{figure=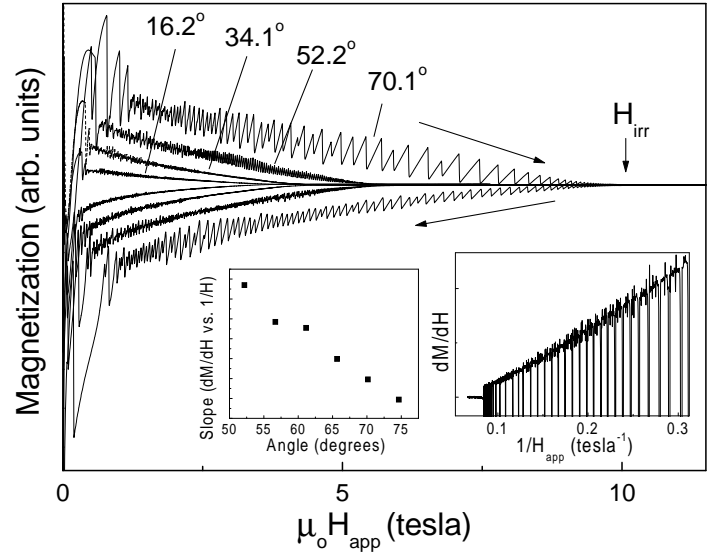,width=100mm}}
\bigskip
\caption{Magnetization vs. magnetic field at 25 mK, for various
angles $\theta$. The direction of the field sweeps is indicated by
arrows. Right inset: d{\bf M}/d{\bf H}$_{app}$ vs. 1/{\bf
H}$_{app}$ shows linear slope above an initial jump.  Left inset:
Slope of d{\bf M}/d{\bf H}$_{app}$ vs. 1/{\bf H}$_{app}$ (right
inset) as a function of $\theta$.} \label{Fig. 1}
\end{figure}


\begin{figure}
\centerline{\epsfig{figure=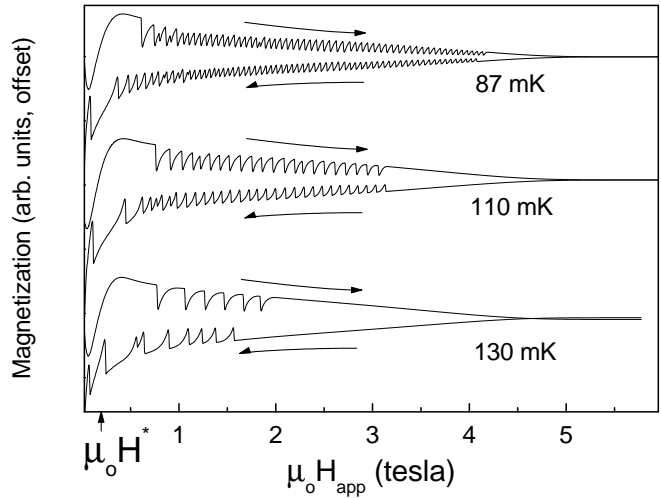,width=100mm}}
\bigskip
\caption{Magnetization vs. applied magnetic field at an angle
$\theta=47.7^o$; the sweep direction is indicated with arrows.
{\bf H*} is the full penetration field in Bean's model [7].}
\label{Fig. 2}
\end{figure}






\end{document}